\documentclass[preprintnumbers,superscriptaddress,showkeys,showpacs,byrevtex]{revtex4}
\usepackage{amsmath,amsfonts,amssymb,amscd,amsxtra,amsthm}
\usepackage{graphicx}
\usepackage{bm}
\usepackage{graphics}
\usepackage{amsmath}
\usepackage{epsfig}
\usepackage{epsf}
\usepackage{epstopdf}
\usepackage{multirow}
\begin{document}
\preprint{CYCU-HEP-14-06}
\preprint{PKNU-NuHaTh-2014-04}
\title{Unpolarized Dihadron Fragmentation Functions in Nonlocal Chiral Quark Model}
\author{Dong-Jing Yang}
\email[E-mail: ]{djyang@std.ntnu.edu.tw}
\affiliation{Department of Physics, National Taiwan Normal University, Taipei 11677, Taiwan}
\affiliation{Institute of Physics, Academia Sinica, Taipei 11529, Taiwan}
\author{Fu-Jiun Jiang}
\email[E-mail: ]{fjjiang@ntnu.edu.tw}
\affiliation{Department of Physics, National Taiwan Normal University, Taipei 11677, Taiwan}
\author{Chian-De Li}
\affiliation{Department of Physics, National Taiwan Normal University, Taipei 11677, Taiwan}
\author{Chung-Wen Kao}
\email[E-mail  (Corresponding Author): ]{cwkao@cycu.edu.tw}
\affiliation{Department of Physics and Center for High Energy Physics, Chung-Yuan Christian University, Chung-Li 32023, Taiwan}
\author{Seung-il Nam}
\email[E-mail: ]{sinam@pknu.ac.kr}
\affiliation{Department of Physics, Pukyong National University (PKNU), Busan 608-737, Republic of Korea}
\date{\today}
\begin{abstract}
We have calculated the unpolarized dihadron fragmentation functions (uDiFFs) of pions and kaons using the nonlocal chiral-quark model (NLChQM) and evolved our results to the transferred momentum scale $Q^2=4\,\mathrm{GeV}^2$ by
the QCD evolution equations. These uDiFFs have also been computed in the Nambu-Jona-Lasinio-jet (NJL-jet) model
for the purpose of comparison. All the calculations have been conducted within the framework of the
Single-Cascade-Algorithm (SCA). We find that there is substantial difference between the results of these two models.
Furthermore, the DiFFs of $u\to \pi^{+}\pi^{-}$ and $g\to \pi^{+}\pi^{-}$ at $Q^2=109\,\mathrm{GeV}^2$ in these two models
are presented in comparison with the parametrizations fitted by the Monte Carlo event generator JETSET.
\end{abstract}
\pacs{12.38.Lg, 13.87.Fh, 12.39.Fe, 14.40.-n, 11.10.Hi.}
\keywords{Kaon and pion fragmentation, flavor SU(3) symmetry breaking, nonlocal chiral-quark model, quark-jet, DGLAP evolution.}
\maketitle
\section{Introduction}
An unpolarized single-hadron fragmentation function $D^{h}_{q}(z,Q^2)$ (uSiFF) describes an unpolarized quark $q$
with the virtuality $Q^2$ to hadronize into a hadron $h$
carrying a fraction of light-cone momentum $z$.
In principle, it can be extracted from experimental data of semi-inclusive processes such as $e^{+}+e^{-}\to h+X$ or $e^-+p\to e+h+X$
with some certain assumptions.
The single-hadron fragmentation functions (SiFFs) include other fragmentation functions
such as the Collins fragmentation function describing the hadronization of a transversely polarized quark.
SiFFs play important roles in the analysis of the scattering processes involving hadrons.
Consequently they have become important subjects in hadronic physics
and have been intensively studied ~\cite{Book}.
Recent development of this subject has been excellently summarized in~\cite{Metz:2016swz}.
\\

N-hadron fragmentation functions are most natural generalization of SiFFs.
They are defined as the overlapping matrices of partonic field operators and N-hadron states.
These functions are essentially non-perturbative objects as the SiFFs. Similar to the SiFFs, N-hadron fragmentation functions are supposed to be factorized from the hard perturbative parts in the hadrons collisions. Furthermore, their evolution with momentum scale $Q^2$ in principle, can be done by applying the perturbative QCD (pQCD).
Similar to the situation of SiFFs~\cite{Armesto:2007dt}, the DGLAP evolution of the N-hadron
fragmentation functions receives medium-induced modification. Therefore one can
compare the N-hadron fragmentation functions in vacuum and the medium and extract
valuable information concerning the medium modification of the multi-particle correlations from jet fragmentation. It
will provide precious knowledge of partonic properties of the dense matter.
This is the main motivation of studying these
objects.\\

In this article, we would only focus on the $N$=2 case. When one analyzes the semi-inclusive processes with two detected hadrons in the final states such as $e^{+}+e^{-}\to h_{1}+h_{2}+X$ or $e^-+p\to e+h_{1}+h_{2}+X$, their cross sections can be written as the convolution of the perturbative
kernel and the unpolarized dihadron fragmentation function $D^{h_1,h_2}_{q}(z_1, z_2, Q^2)$ (uDiFF) which is the probability of a quark $q$ fragmenting into two hadrons $h_1$ and $h_2$ with the light-cone momentum fractions $z_1$ and $z_2$, respectively~\cite{Konishi78}. The QCD evolution equations of uDiFFs have been intensively investigated
~\cite{Vendramin81,Sukhatme80,deFlorian04,Majumder04,Majumder05}.\\

\begin{figure}[t]
\begin{tabular}{c}
\includegraphics[width=8.5cm]{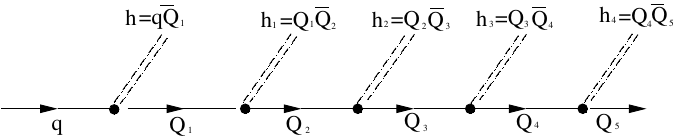}
\end{tabular}
\caption{Quark fragmentation cascade process. The Single-Cascade-Algorithm (SCA) is based on this kind of picture. }
\label{cascade}
\end{figure}

To investigate those SiFFs and DiFFs one needs rely on effective QCD models,
because the fragmentation functions are defined in Minkowski
space such that the usual lattice QCD techniques are not applicable.
Furthermore, the hadronization of outgoing quark and gluon jets is essentially a nonperturbative phenomenon governed
by the long-distance physics. Hence the perturbative QCD is hardly useful except for their QCD evolutions.
It is particularly interesting to apply the chiral models to study those fragmentation functions
since the chiral dynamics is an important nonperturbative QCD effect which plays a dominant role in the phenomenology
of QCD in the low energy regime.\\

There are several chiral models which have been adopted to
study the fragmentation functions.
For example, the chiral quark model of Manohar and Georgi ~\cite{MG} has been used to calculate the unpolarized fragmentation functions ~\cite{Collins93} and the Collins fragmentation function
~\cite{Bacchetta:2002tk}. DiFFs have been calculated in the spectator quark model in Ref.~\cite{Bacchetta:2006un}.
The NJL-jet model has also been developed to compute
both of uSiFFs and uDiFFs~\cite{Matevosyan:2010hh,Matevosyan:2011ey,Matevosyan:2011vj,Casey:2012ux,Casey:2012hg}.
We have adopted the nonlocal chiral-quark model (NLChQM) to study the uSiFFs of the pions in Ref.~\cite {Nam:2011hg}.
Our result has been extended to the uSiFFs of kaons later~\cite{Nam:2012af}.
Furthermore we have included the quark-jet contribution to SiFFs of the pions and the kaons in Ref.~\cite{Yang:2013cza}.
It has been found that the two models produce very different results of the uSiFFs of the pions and kaons.
In particular they own very distinct patterns of the SU(3) breaking effect.
Recently we have used the uSiFFs of both models to study the charged meson multiplicities in HERMES Semi-Inclusive Deep
Inelastic Scattering (SIDIS) data in Ref.~\cite{HERMES} and obtain rather different interpretations of the HERMES data~\cite{Yang:2015avi}.
Consequently, we would like to extend our investigation to uDiFFs $D^{h_{1},h_{2}}_{q}(z_1,z_2,Q^2)$ which are less known empirically, compared with the other fragmentation functions. \\

This article is organized as follows: We briefly review the process of computing uSiFFs in the nonlocal
chiral-quark model in Sec.~II.
In Sec.~III we describe how to obtain uDiFFs in the NLChQM and the NJL-jet model within the framework of the Single-Cascade-Algorithm (SCA).
We present and discuss our results which have been evolved to $Q^2=4\,\mathrm{GeV}^2$ in Sec.~IV.
Finally, we make our conclusion in Sec.~V. In the appendix, we figure out the result of Ref.~\cite{Casey:2012ux} is actually incorrect, and consequently, the result of Ref.~\cite{Casey:2012ux} is also incorrect.

\section{Unpolarized Single-Hadron Fragmentation Function in NLChQM}
\begin{figure}[b]
\begin{tabular}{ccc}
\includegraphics[width=5.2cm]{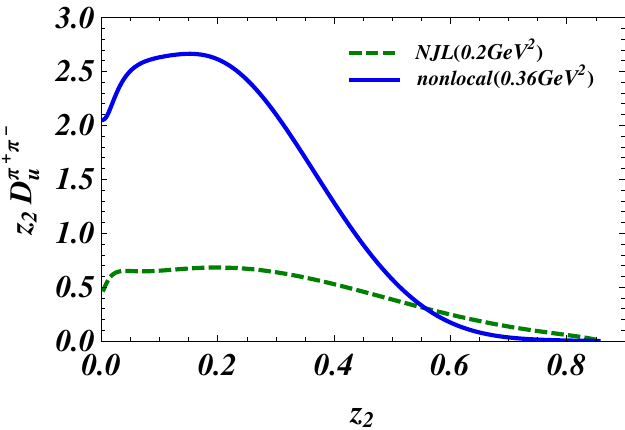}
\includegraphics[width=5.2cm]{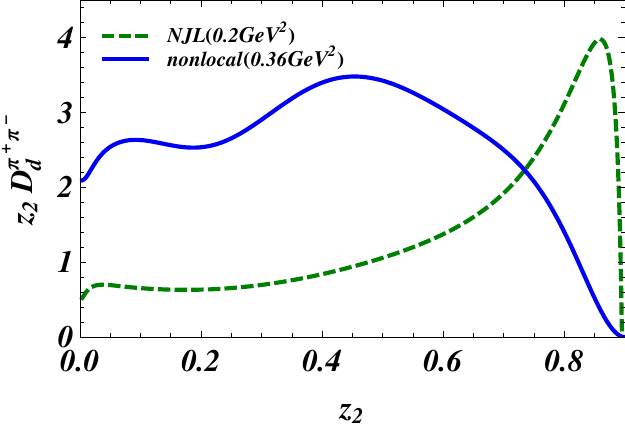}
\includegraphics[width=5.2cm]{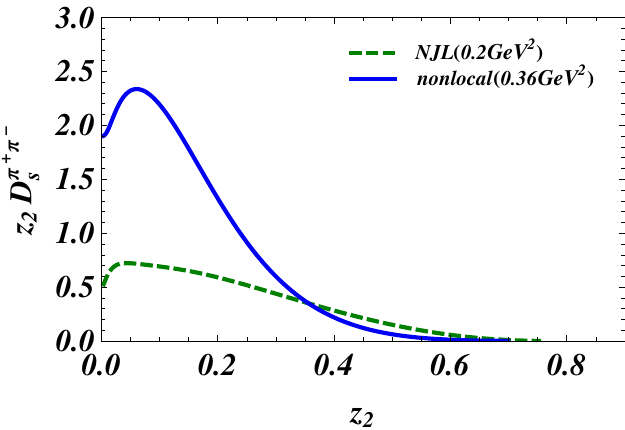}
\end{tabular}
\begin{tabular}{ccc}
\includegraphics[width=5.2cm]{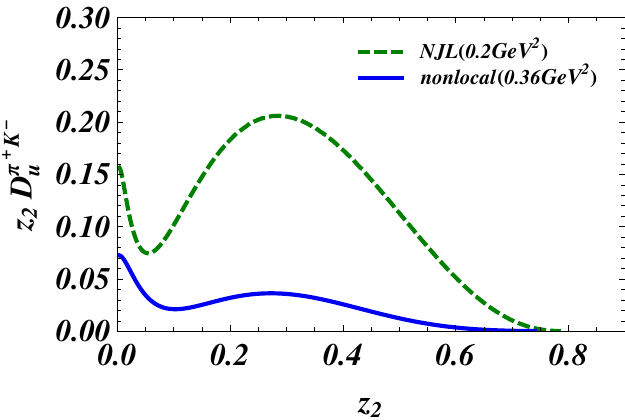}
\includegraphics[width=5.2cm]{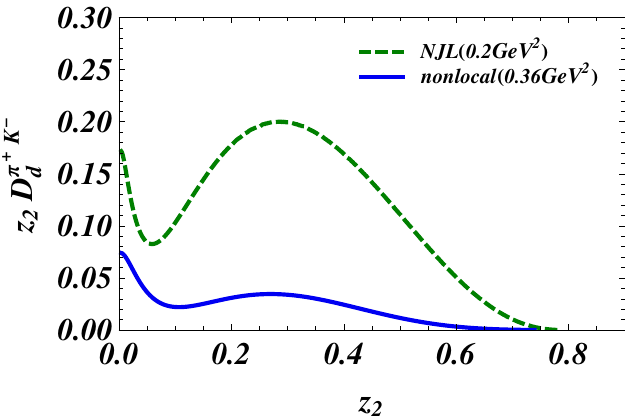}
\includegraphics[width=5.2cm]{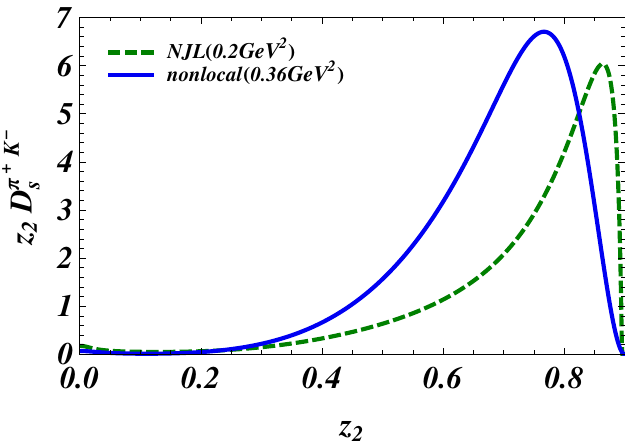}
\end{tabular}
\begin{tabular}{ccc}
\includegraphics[width=5.2cm]{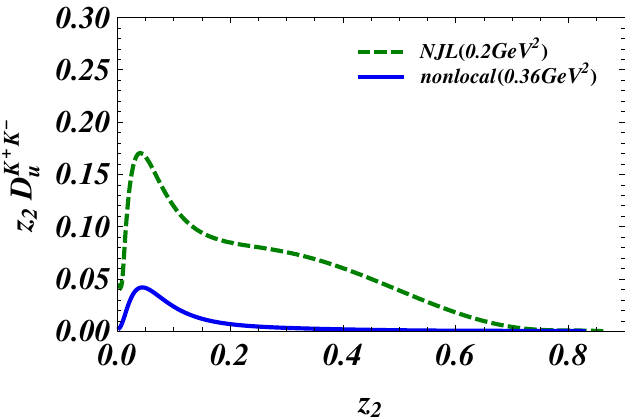}
\includegraphics[width=5.2cm]{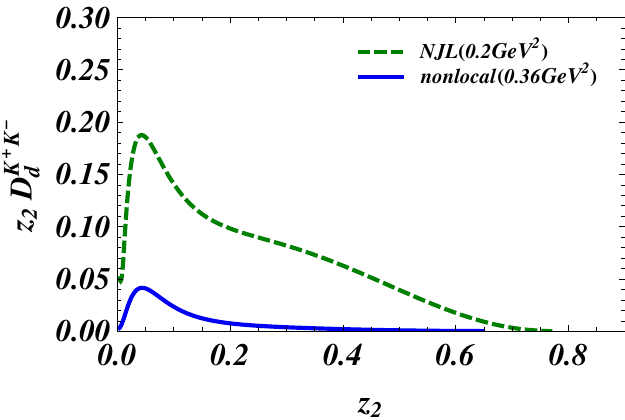}
\includegraphics[width=5.2cm]{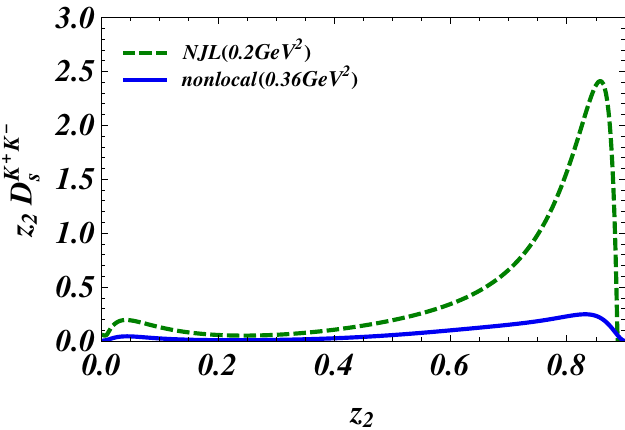}
\end{tabular}
\caption{$z_2 D^{h_1,h_2}_{q}(z_1,z_2)$ with $z_{1}=0.1$ for
(1) $(q,h_1,h_2)=(u,\pi^{+},\pi^{-})$ (left of the top row), (2)$(d,\pi^{+},\pi^{-})$
(middle of the top row), (3) $(s,\pi^{+},\pi^{-})$ (right of the top row),
(4) $(u,\pi^{+},K^{-})$(left of the middle row), (5) $(d,\pi^{+},K^{-})$
(middle of the middle row), (6) $(s,\pi^{+},K^{-})$ (right of the middle row),
(7) $(u,K^{+},K^{-})$ (left of the bottom row), (8) $(d,K^{+},K^{-})$
(middle of the bottom row), (9) $(s,K^{+},K^{-})$ (right of the bottom row).
The dashed and solid lines denote the results of the NJL-jet model and the nonlocal chiral quark model, respectively.
The range of $z_2$ is from zero to 0.9. }
\label{A}
\end{figure}
In this section, we briefly explain how to compute uSiFFs in the NLChQM. The details of the derivation can be found in Refs.~\cite{Nam:2011hg,Nam:2012af,Yang:2013cza}. The NLChQM is motivated from the Dilute Instanton-Liquid Model (DILM)~\cite{Diakonov:1985eg,Shuryak:1981ff,Diakonov:1983hh,Diakonov:2002fq,Schafer:1996wv} where
the quark-instanton interactions induced by the dilute instanton ensemble is to generate
the nonperturbative QCD effects. Although DILM is defined in Euclidean space because the
(anti)instantons are only well defined there, several works have
replaced Euclidean metric for the (anti)instanton
effective chiral action with the one of Minkowski space
~\cite{Dorokhov:1991nj,Nam:2006sx,Nam:2006au,Praszalowicz:2001pi}.
The model constructed by this way
is called the nonlocal chiral-quark model (NLChQM) because the interactions between the chiral fields and the constituent quarks are nonlocal.
We reach a concise expression for the elementary uSiFF $\hat{d}^{h}_{q}$ describing the fragmentation process
$q(k)\to h(p)+Q(k-p)$ from the NLChQM as follows:


\begin{figure}[t]
\begin{tabular}{ccc}
\includegraphics[width=5.2cm]{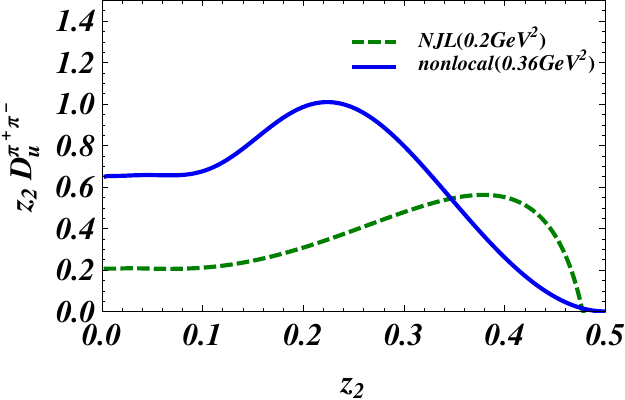}
\includegraphics[width=5.2cm]{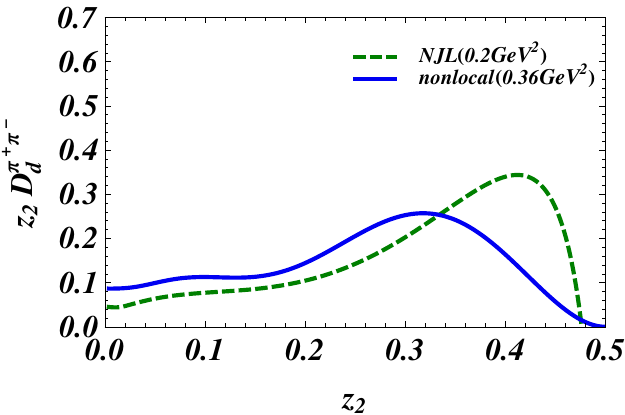}
\includegraphics[width=5.2cm]{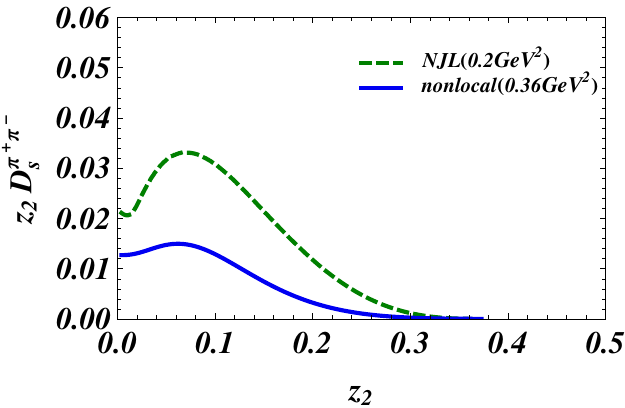}
\end{tabular}
\begin{tabular}{ccc}
\includegraphics[width=5.2cm]{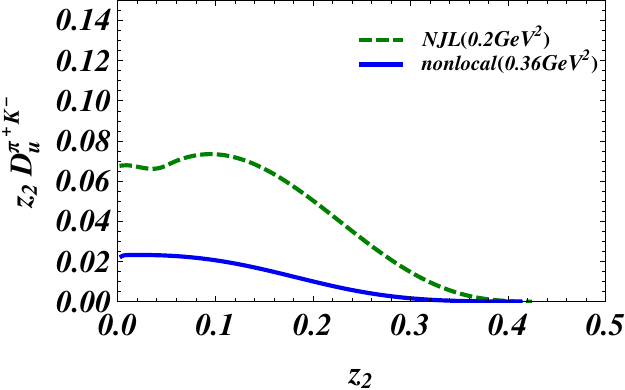}
\includegraphics[width=5.2cm]{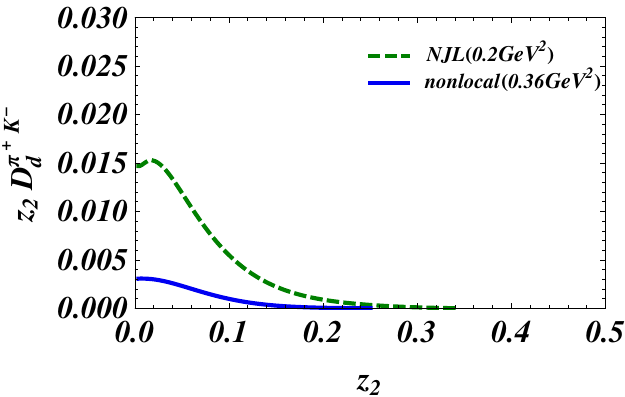}
\includegraphics[width=5.2cm]{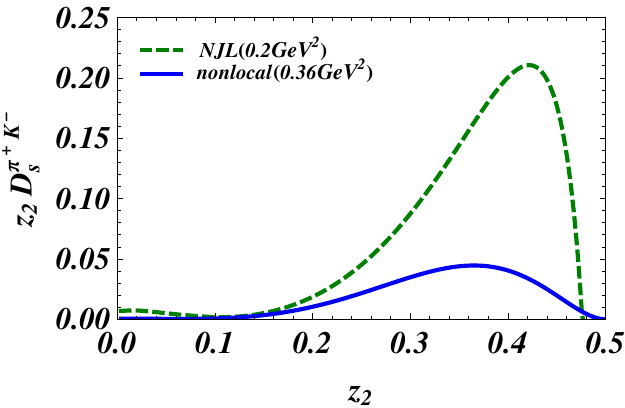}
\end{tabular}
\begin{tabular}{ccc}
\includegraphics[width=5.2cm]{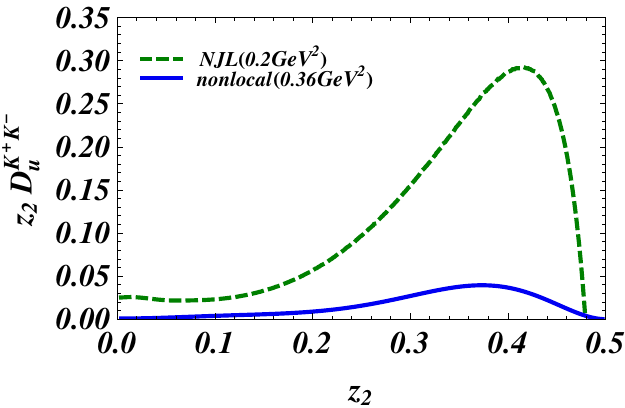}
\includegraphics[width=5.2cm]{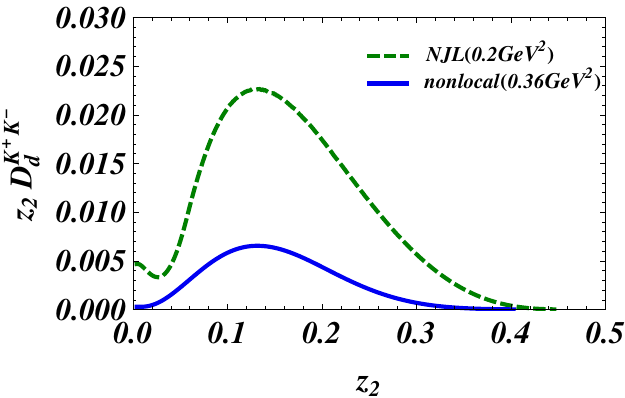}
\includegraphics[width=5.2cm]{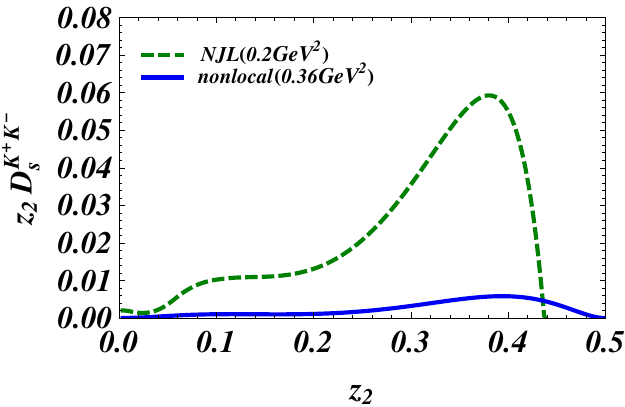}
\end{tabular}
\caption{$z_2 D^{h_1,h_2}_{q}(z_1,z_2)$ with $z_{1}=0.5$ for
(1) $(q,h_1,h_2)=(u,\pi^{+},\pi^{-})$ (left of the top row), (2) $(d,\pi^{+},\pi^{-})$
(middle of the top row), (3) $(s,\pi^{+},\pi^{-})$ (right of the top row),
(4) $(u,,\pi^{+},K^{-})$(left of the middle row), (5) $(d,\pi^{+},K^{-})$
(middle of the middle row), (6) $(s,\pi^{+},K^{-})$ (right of the middle row),
(7) $(u,K^{+},K^{-})$ (left of the bottom row), (8) $(d,K^{+},K^{-})$
(middle of the bottom row), (9) $(s,K^{+},K^{-})$ (right of the bottom row). The dashed and solid lines denote the results of the NJL-jet model and the nonlocal chiral
quark model, respectively. The range of $z_2$ is from zero to 0.5.}
\label{B}
\end{figure}
\begin{equation}
d^{h}_{q}(z,\bm{k}^2_T,\Lambda)=\frac{\mathcal{C}^h_{q}}
{8\pi^3}\frac{M_kM_{r}}{2F^2_h}
\frac{z\left[z^2\bm{k}^2_T+[(z-1)\bar{M}_q+\bar{M}_{Q}]^2\right]}
{[z^2\bm{k}^2_T+z(z-1)\bar{M}^2_q+z\bar{M}^2_{Q}+(1-z)m^2_h]^2}.
\label{eq:DDDDD}
\end{equation}
Here the following notations are used:
$F_ h$ stands for the weak-decay constant for the pseudo-scalar (PS) meson $h$ whose flavor content is $\bar{q}Q$.
$\mathcal{C}^h_{q}$ indicates the flavor factor for the corresponding fragmentation processes listed in
Table~(\ref{TABLE0}).
$\bar{M}_q$ is defined as
$\bar{M}_q\equiv m_q+M_0$. Here $m_q$ is the current quark mass for the light quarks: $m_u=m_d=5$ MeV and $m_s=150$ MeV.
The value of $M_0$ can be fixed
self-consistently within the instanton model~\cite{Diakonov:1985eg,Shuryak:1981ff,Diakonov:1983hh,Schafer:1996wv,Musakhanov:1998wp,Musakhanov:2002vu,Diakonov:2002fq,
Nam:2007gf,Nam:2010pt} with the phenomenological (anti)instanton parameters $\bar{\rho}\approx1/3$ fm and $\bar{R}\approx1$ fm.
This leads to $M_0\approx350$ MeV.
In addition, $M_k$ and $M_{r}$ appearing in Eq.~(\ref{eq:DDDDD}) are the momentum-dependent quark masses manifesting the nonlocal quark-PS meson interactions:
\begin{equation}
\label{eq:MASS}
M_k=\frac{M_0[2\Lambda^2z(1-z)]^2}
{[z^2\bm{k}^2_T+z(z-1)(2\Lambda^2-\delta^2)+z\bar{M}^2_{Q}+(1-z)m^2_h]^2},
\,\,\,\,
M_{r}=\frac{M_0(2\Lambda^2)^2}{(2\Lambda^2-\bar{M}^2_{Q})^2}.
\end{equation}
As explained in Ref.~\cite{Nam:2011hg}, a free and finite-valued parameter $\delta$ has been
introduced in the denominator to avoid the unphysical singularities.
$\Lambda$ is the cut-off scale implied in this model.
Notice that the singularities arise in the vicinity of $(z,\bm{k}_T)=0$. The elementary uSiFF can
be evaluated further by integrating Eq.~(\ref{eq:DDDDD}) over $\bm{k}_T$:
\begin{equation}
\label{eq:FRAGINT}
d^{h}_{q}(z,\Lambda)=2\pi z^2\int^\infty_0 d^{h}_{q}(z,\bm{k}^2_T,\Lambda)
\,\bm{k}_T\,d\bm{k}_T.
\end{equation}
Note that both $M_{k}$ and $M_{r}$ depend on $\bm{k}_{T}$ such that the integration in Eq. ~(\ref{eq:FRAGINT})
converges.
\begin{table}[h]
\begin{tabular}{c|ccccccc}
$\mathcal{C}^{h}_q$&$\pi^0$&$\pi^+$&$\pi^-$&$K^0$&$\bar{K^{0}}$&$K^+$&$K^-$\\
\hline
$u$&$1/2$&$1$&$0$&$0$&$0$&$1$&$0$\\
$d$&$1/2$&$0$&$1$&$1$&$0$&$0$&$0$\\
$s$&$0$&$0$&$0$&$0$&$1$&$0$&$1$\\
$\bar{u}$&$1/2$&$0$&$1$&$0$&$0$&$0$&$1$\\
$\bar{d}$&$1/2$&$1$&$0$&$0$&$1$&$0$&$0$\\
$\bar{s}$&$0$&$0$&$0$&$1$&$0$&$1$&$0$\\
\end{tabular}
\caption{Flavor factors in Eq.~(\ref{eq:DDDDD}).}
\label{TABLE0}
\end{table}
To include the quark-jet contribution
we have followed the approach developed in ~\cite{Matevosyan:2010hh, Matevosyan:2011ey,Matevosyan:2011vj}.
First, the elementary uSiFF $\hat{d}^{h}_{q}(z)$ is re-defined as follows:
\begin{equation}
\sum_{h}\int\hat{d}^{h}_{q}(z)dz=\sum_{Q}\int\hat{d}^{Q}_{q}(z)dz=1,
\end{equation}
where the complementary uSiFF $\hat{d}^{Q}_{q}(z)$ is given by
\begin{equation}
\hat{d}^{Q}_{q}(z)=\hat{d}^{h}_{q}(1-z),\,\,\,\,\,\, h=q\bar{Q}.
\end{equation}
The full uSiFF $D^{h}_{q}(z)$ should satisfy the following integral equation:
\begin{equation}
D^{h}_{q}(z)dz=\hat{d}^{h}_{q}(z)dz+\sum_{Q}\int^{1}_{z}dy\,
\hat{d}^{Q}_{q}(y)D^{h}_{Q}\left(\frac{z}{y}\right)\frac{dz}{y}.
\label{multijet}
\end{equation}
Here $D^{h}_{q}(z)dz$ is the probability for a quark $q$ to emit a hadron $h$
which carries the light-cone momentum fraction from $z$ to $z+dz$.
$\hat{d}^{Q}_{q}(y)dy$ is the probability for a quark $q$ to emit a hadron with flavor composition $q\bar{Q}$ at one step and the final quark becomes
$Q$ with the light-cone momentum fraction from $y$ to $y+dy$.
Eq.~(\ref{multijet}) actually describes a fragmentation cascade process of hadron
emissions of a single quark depicted in Fig.~(\ref{cascade}).

One can either solve the coupled integral equations in Eq.~(\ref{multijet}) by iteration or by the Monte Carlo (MC) method developed in ~\cite{Matevosyan:2011ey}.
The MC method is to simulate the
fragmentation cascade of a quark through $N_\mathrm{tot}$ times, and at each time the fragmentation cascade stops after the quark emits
$N_\mathrm{links}$ hadrons. $D^{h}_{q}(z)$ is then extracted through the average number of type $h$ hadron with light-cone momentum
fraction $z$ to $z+\Delta z$, $N^{h}_{q}(z,z+\Delta z)$:
\begin{equation}
D^{h}_{q}(z)\Delta z
=\frac{1}{N_\mathrm{tot}}\sum_{N_\mathrm{tot}}N^{h}_{q}(z,z+\Delta z).
\end{equation}
The value of $D^{h}_{q}(z)\Delta z$ becomes insensitive to the value of $N_\mathrm{tot}$ and
$N_\mathrm{links}$, when $N_\mathrm{tot}$ and $N_\mathrm{links}$ are large enough, implying that the result of the MC simulation converges to the solution
of Eq.~(\ref{multijet}).
Once the forms of $\hat{d}^{h}_{q}$ are given by the certain models then one can
derive the associated uSiFF $D^{h}_{q}$.
The result of $D^{h}_{q}$ in the NLChQM has been presented in ~\cite{Yang:2013cza}.
\begin{figure}[t]
\begin{tabular}{ccc}
\includegraphics[width=5.2cm]{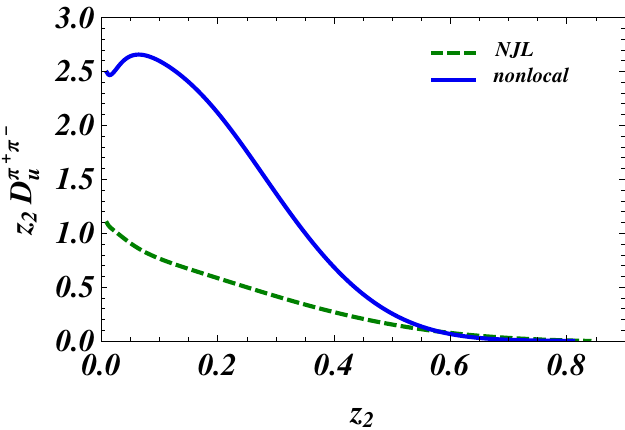}
\includegraphics[width=5.2cm]{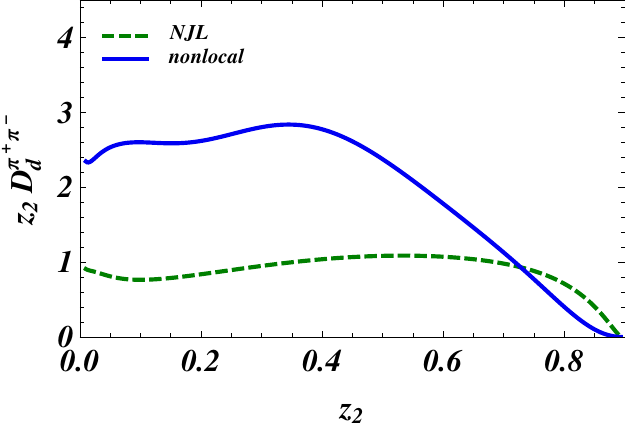}
\includegraphics[width=5.2cm]{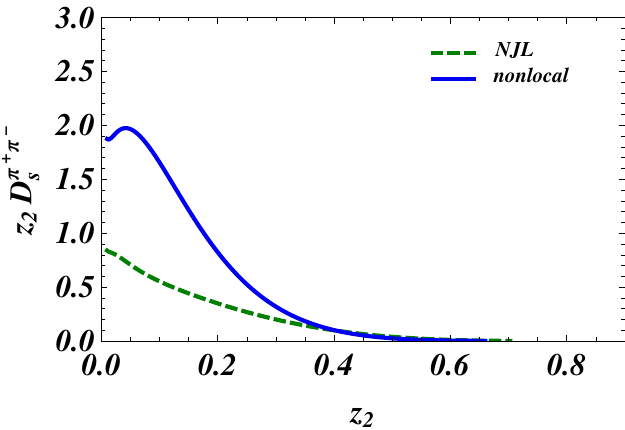}
\end{tabular}
\begin{tabular}{ccc}
\includegraphics[width=5.2cm]{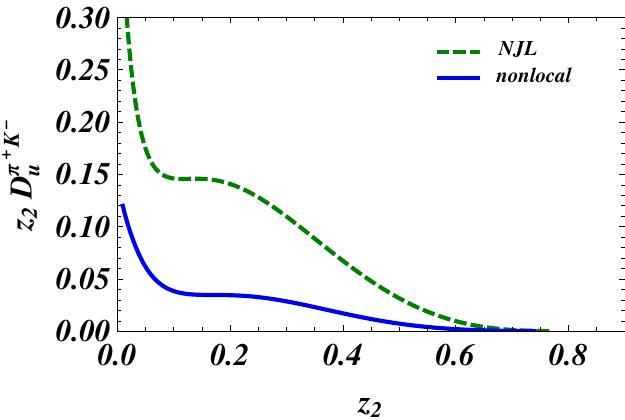}
\includegraphics[width=5.2cm]{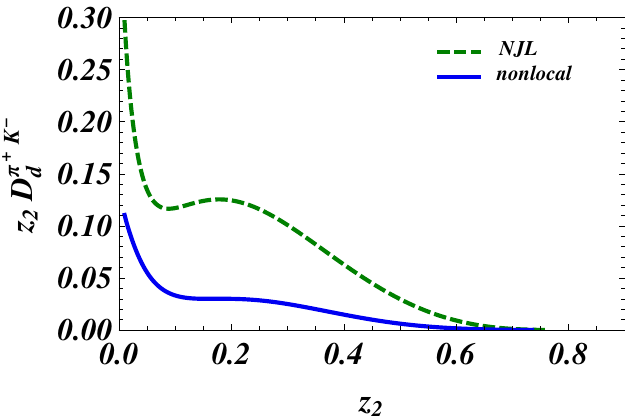}
\includegraphics[width=5.2cm]{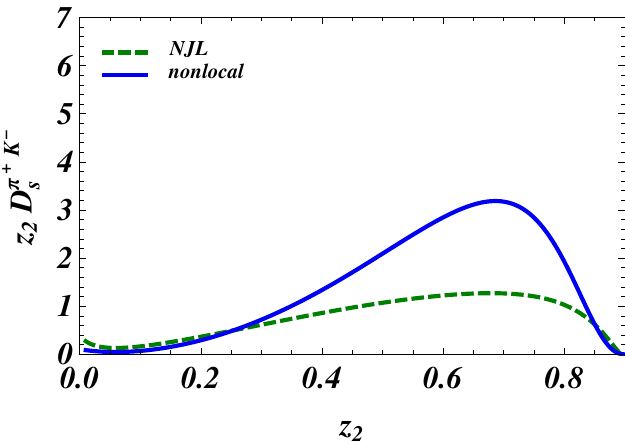}
\end{tabular}
\begin{tabular}{ccc}
\includegraphics[width=5.2cm]{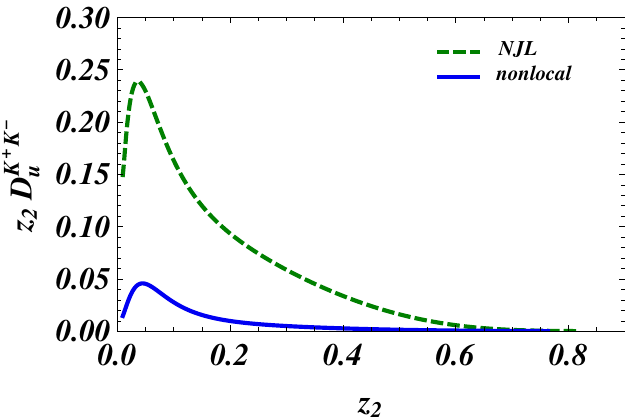}
\includegraphics[width=5.2cm]{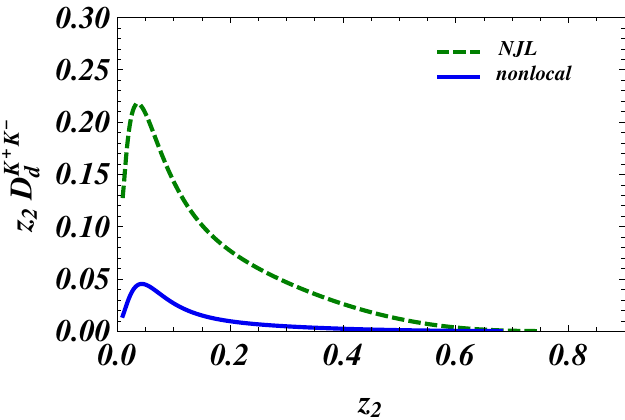}
\includegraphics[width=5.2cm]{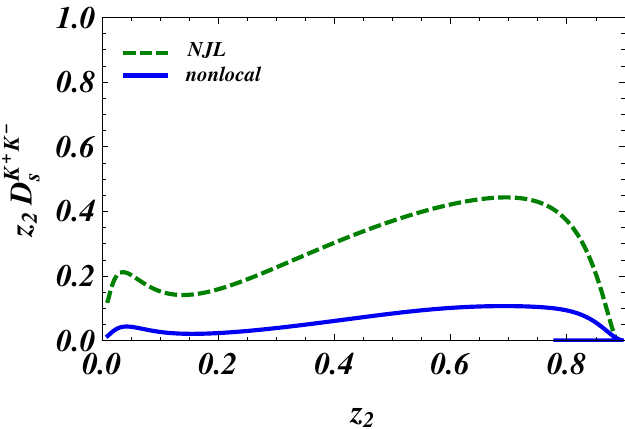}
\end{tabular}
\caption{$z_2 D^{h_1,h_2}_{q}(z_1,z_2)$ with $z_{1}=0.1$ and $Q^2=4\,\mathrm{GeV}^2$ for
(1) $(q,h_1,h_2)=(u,\pi^{+},\pi^{-})$ (left of the top row), (2) $(d,\pi^{+},\pi^{-})$
 (middle of the top row), (3) $(s,\pi^{+},\pi^{-})$ (right of the top row),
(4) $(u,\pi^{+},K^{-})$ (left of the middle row), (5) $(d,\pi^{+},K^{-})$
(middle of the middle row), (6) $(s,\pi^{+},K^{-})$ (right of the middle row),
(7) $(u,K^{+},K^{-})$ (left of the bottom row), (8) $(d,K^{+},K^{-})$
 (middle of the bottom row), (9) $(s,K^{+},K^{-})$ (right of the bottom row).
The dashed and solid lines denote the results of the NJL model and the nonlocal chiral quark model, respectively.
The range of $z_2$ is from zero to 0.9.}
\label{C}
\end{figure}

\section{Unpolarized Dihadron Fragmentation Functions in NLChQM}
In this section we describe how to obtain the uDiFFs within our model.
Similar to the case of uSiFFs, there are also
two ways to reach the goal. The first method is the Monte Carlo (MC) simulation.
We apply the MC method to simulate the
fragmentation cascade of a quark via $N_\mathrm{tot}$ times and each time the fragmentation cascade stops
after the quark emits $N_\mathrm{links}$ hadrons.
The fragmentation function $D^{h_1,h_2}_{q}(z)$ is then extracted through
the average number of type $h_1$ hadron with light-cone momentum fraction from $z_1$ to $z_1+\Delta z_1$ together with
the type $h_2$ hadron with light-cone momentum fraction from $z_2$ to $z_2+\Delta z_2$,
$N^{h_1,h_2}_{q}(z_1,z_1+\Delta z_1,z_2+\Delta z_2)$ as follows,
\begin{equation}
D^{h_1,h_2}_{q}(z_1,z_2)\Delta z_1\Delta z_2
=\frac{1}{N_\mathrm{tot}}\sum_{N_\mathrm{tot}}N^{h_1,h_2}_{q}(z_1,z_1+\Delta z_1;z_2,z_2+\Delta z_2).
\end{equation}

The second way is to relate the uDiFFs with the uSiFF $D^{h}_{q}(z)$ and the elementary uSiFF $\hat{d}^{h}_{q}(z)$
in according to the following equation \cite{Field} :
\begin{equation}
D^{h_{1},h_{2}}_{q}(z_{1},z_{2})=\delta_{aq}\hat{d}^{h_{1}}_{q}(z_{1})\frac{D^{h_{2}}_{q_{1}}\left(\frac{z_{2}}{1-z_{1}}\right)}{1-z_{1}}+
\delta_{bq}\hat{d}^{h_{2}}_{q}(z_{2})\frac{D^{h_{1}}_{q_{2}}\left(\frac{z_{1}}{1-z_{2}}\right)}{1-z_{2}}
+\sum_{Q}\int^{1}_{z_{1}+z_{2}}\frac{d\eta}{\eta^2}\hat{d}^{Q}_{q}(\eta)D^{h_{1},h_{2}}_{Q}\left(\frac{z_{1}}{\eta},\frac{z_2}{\eta}\right).
\label{Dihadron1}
\end{equation}
Here the flavor component of the emitted hadrons $h_1$ and $h_2$ are
$h_{1}=(a\bar{q}_{1})$ and $h_{2}=(b\bar{q}_{2})$, respectively.
If $q$ is neither $a$ nor $b$
then $D^{h_1,h_2}_{q}$ is called the disfavored uDiFF. Otherwise it is called the favored uDiFF.
The first term stands for the situation that $h_1$ is the first emitted hadron in the decay cascade of the quark $q$.
Similarly the second term denotes the situation that $h_2$ is
the first emitted hadron. The third term represents the situation that the first emitted hadron is neither $h_{1}$ nor $h_{2}$. To simplify the equation we make change of valuables
as $\xi_{1}=z_{1}/\eta$ and $\xi_{2}=z_2/\eta$:
\begin{eqnarray}
D^{h_{1},h_{2}}_{q}(z_{1},z_{2})&=&\delta_{qa}\hat{d}^{h_{1}}_{q}(z_{1})\frac{D^{h_{2}}_{q_{1}}\left(\frac{z_{2}}{1-z_{1}}\right)}{1-z_{1}}+
\delta_{qb}\hat{d}^{h_{2}}_{q}(z_{2})\frac{D^{h_{1}}_{q_{2}}\left(\frac{z_{1}}{1-z_{2}}\right)}{1-z_{2}} \nonumber \\
&+&\sum_{Q}\int^{\frac{z_1}{z_{1}+z_{2}}}_{z_{1}}d\xi_{1}\int^{\frac{z_1}{z_1+z_2}}_{z_2}d\xi_{2}\delta(z_2\xi_{1}-z_{1}\xi_{2})
\hat{d}^{Q}_{q}(z_{1}/\xi_{1})D^{h_{1},h_{2}}_{Q}(\xi_{1},\xi_{2}).
\label{Dihadron2}
\end{eqnarray}


Adopting the $\hat{d}^{h}_{q}(z)$ in Eq.~(\ref{eq:FRAGINT}) from the NLChQM, we first solve Eq. ~(\ref{multijet}) to
obtain the uSiFFs $D^{h}_{q}$. Then we employ the technique of iteration to solve
Eq.~(\ref{Dihadron1}). We find that the results agree with the one obtained by the MC method excellently.
Similarly we also take $\hat{d}^{h}_{q}(z)$ of the NJL-jet model to calculate the uDiFFs. The result of solving
Eq.~(\ref{Dihadron1}) by the iteration method also agree with the result of the MC simulation.

Notice the uDiFFs of the NJL-jet model have been
presented in ~\cite{Casey:2012ux}.
However, we notice that our results of the NJL-jet model are different from those in Ref.~\cite{Casey:2012ux}.
The reason of this disagreement is explained in the appendix.
Through this article whenever the NJL-jet model results are mentioned,
they are referred to the ones obtained by our calculation.

The NLChQM results of $z_{2}D^{h_{1},h_{2}}_{q}(z_1,z_2)$ are presented in Fig.~(\ref{A}) for $z_1=0.1$ and in Fig.~(\ref{B}) for $z_1$=0.5. The results of
the NJL-jet model are also presented in the same figures for comparison.
We only present the cases of $(h_1,h_2)=(\pi^{+}\pi^{-})$, $(\pi^{+}K^{-})$, and $(K^{+}K^{-})$
here. Among the nine uDiFFs we present here, only three of them, $D^{\pi^{+},\pi^{-}}_{s}$, $D^{\pi^{+},K^{-}}_{d}$
and $D^{K^{+},K^{-}}_{d}$ are disfavored ones. The other six uDiFFs are all the favored ones.

In general, we find that the curves of the results based on NLChQM are substantially different from the curves
based on the results of NJL-jet model both in the magnitude and the shapes.
For $z_1=0.1$, the uDiFFs of the $\pi^{+}\pi^{-}$ pair in the NLChQM are twice larger than the ones in the NJL-jet model
in the low $z_2$ region. However,
for the $\pi^{+}K^{-}$ pair, the NLChQM results are about one third of the NJL-jet results except for $s\to \pi^{+}K^{-}$
where the results of the two models are about
the same magnitude. Moreover, the NLChQM results of uDiFFs of the $K^{+}K^{-}$ pair are much smaller than
the corresponding ones in the NJL-jet model.

The difference between the results from NLChQM and the NJL-jet model is reduced as $z_1$ increases. In general the magnitudes of the curves are all reduced
compared with the case of $z_1=0.1$.
The relation between the results in the two models are also changed in some cases.
For example, the NLChQM results of the DiFFs
for $ s\to \pi^{+}\pi^{-}$ and $ s\to \pi^{+}K^{-}$ become smaller than the NJL-jet result at $z_1=0.5$, but
at $z_1=0.1$ the NLChQM results is larger than the ones in the NJL-jet model.

Notice that the results of the two models in Figs.~(\ref{A}) and (\ref{B}) actually correspond to the different $Q^2$ values.
While the NJL-jet model results are set at
$Q^2=0.2\,\mathrm{GeV}^2$, the scale of the NLChQM results is $Q^2=0.36\,\mathrm{GeV}^2$ ~\cite{Yang:2013cza}.
In the next section, we will apply the QCD evolution to the both results and compare
them at the same $Q^2$ value.
\section{QCD Evolution of Dihadron Fragmentation Functions}
\begin{figure}[t]
\begin{tabular}{ccc}
\includegraphics[width=5.2cm]{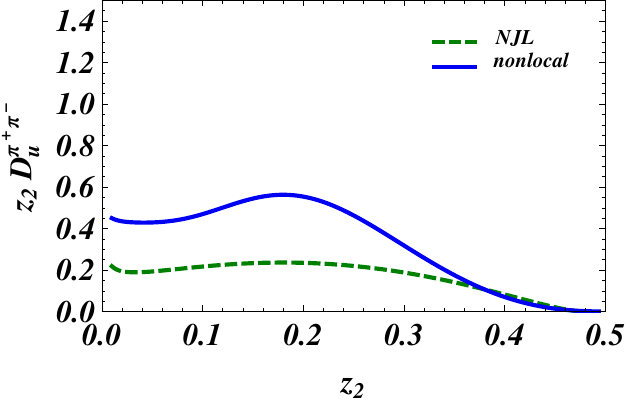}
\includegraphics[width=5.2cm]{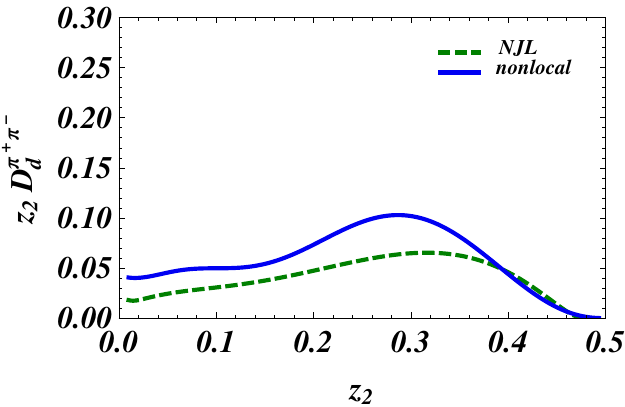}
\includegraphics[width=5.2cm]{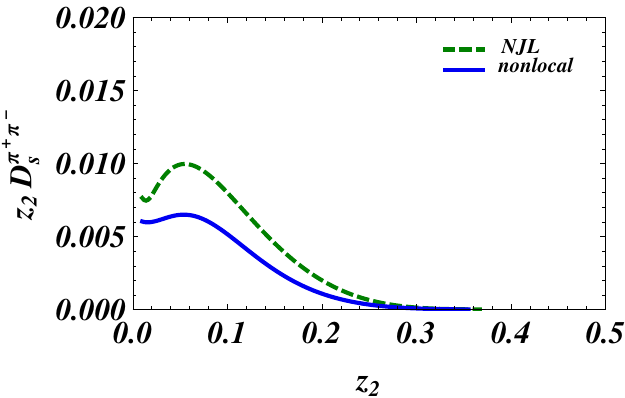}
\end{tabular}
\begin{tabular}{ccc}
\includegraphics[width=5.2cm]{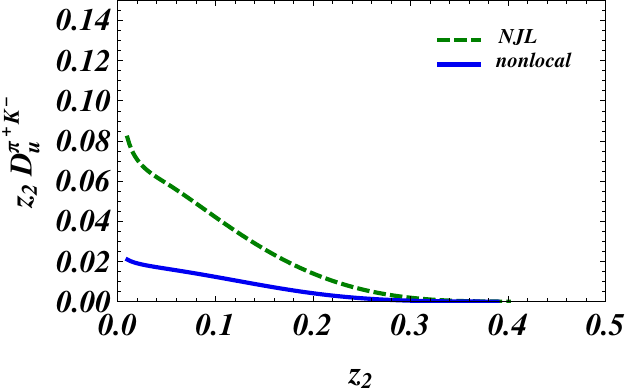}
\includegraphics[width=5.2cm]{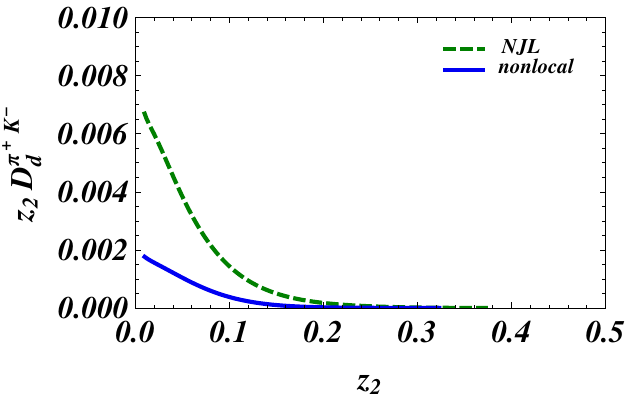}
\includegraphics[width=5.2cm]{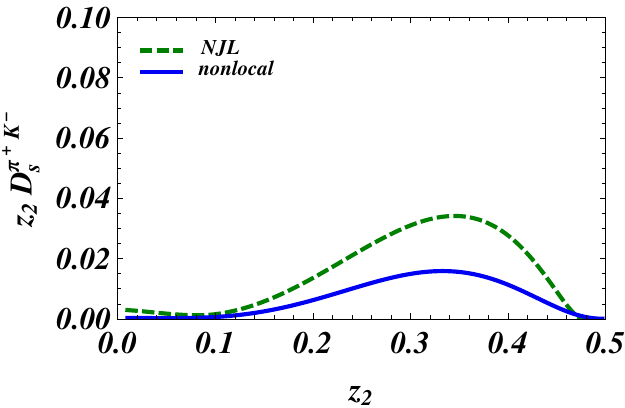}
\end{tabular}
\begin{tabular}{ccc}
\includegraphics[width=5.2cm]{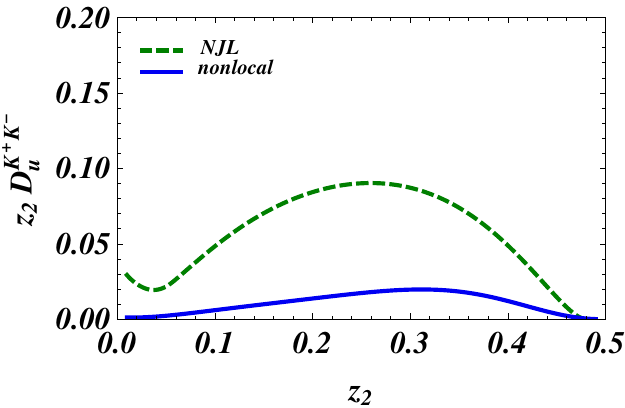}
\includegraphics[width=5.2cm]{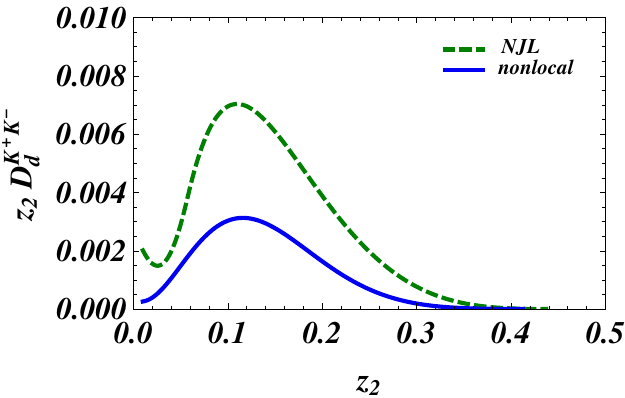}
\includegraphics[width=5.2cm]{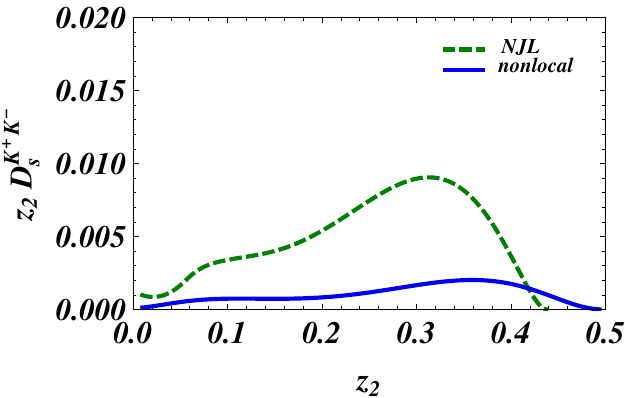}
\end{tabular}
\caption{$z_2 D^{h_1,h_2}_{q}(z_1,z_2)$ with $z_{1}=0.5$ and $Q^2=4\,\mathrm{GeV}^2$ for
(1) $(q,h_1,h_2)=(u,\pi^{+},\pi^{-})$ (left of the top row), (2) $(d,\pi^{+},\pi^{-})$
(middle of the top row), (3) $(s,\pi^{+},\pi^{-})$ (right of the top row),
(4) $(u,\pi^{+},K^{-})$ (left of the middle row), (5) $(d,\pi^{+},K^{-})$
(middle of the middle row), (6) $(s,\pi^{+},K^{-})$ (right of the middle row)
(7) $(u,K^{+},K^{-})$ (left of the bottom row), (8) $(d,K^{+},K^{-})$
(middle of the bottom row), (9) $(s,K^{+},K^{-})$ (right of the bottom row).
The dashed and solid lines denote the results of the NJL-jet model and the nonlocal chiral quark model respectively. The range of $z_2$ is from zero to 0.5.
}
\label{D}
\end{figure}

The QCD evolution equations of uDiFFs
have been derived in Ref.~\cite{deFlorian04}.
The QCD evolution equations of the uDiFFs are as follow:
\begin{eqnarray}
\frac{d}{d\ln Q^2}D_{i}^{h_1,h_2}(z_1,z_2,Q^2)&=&\frac{\alpha_s(Q^2)}{2\pi}\int^{1}_{z_1+z_2}\frac{du}{u^2}D^{h_1,h_2}_{j}\left(\frac{z_1}{u},\frac{z_2}{u},Q^2 \right)
P_{ji}(u) \nonumber \\
&+&\frac{\alpha_{s}(Q^2)}{2\pi}\int^{1-z_2}_{z_1}\frac{du}{u(1-u)}D^{h_{1}}_{j}\left(\frac{z_1}{u},Q^2 \right)
D^{h_{2}}_{k}\left(\frac{z_2}{u},Q^2 \right)\hat{P}_{kj}(u),
\end{eqnarray}
where the Latin indices $i,j,k$ can be a quark, antiquark or gluon. Similar to our
previous work~\cite{Yang:2013cza}, we assume uDiFFs for the gluon are identical
zero at the initial $Q^2_0$ value.
The value of the initial $Q_0^2$ is
set to be $0.36\,\mathrm{GeV}^2$ for the NLChQM~\cite{Yang:2013cza} and $0.2 \,\mathrm{GeV}^2$ for
the NJL-jet model.
Our results of uDiFFs at $Q^2=4\,\mathrm{GeV}^2$ are
presented in Figs.~(\ref{C}) for $z_1=0.1$ and (\ref{D}) for $z_1=0.5$.  We also present
the results of the NJL-jet model in the same figures for comparison. Note that our NJL-jet model result is different
with the ones in Ref.~\cite{Casey:2012hg}. It is because their result is the result of QCD evolution of the result of Ref.~\cite{Casey:2012ux} which is
incorrect. We will explain it in detail in the appendix.

In general, we find that the QCD evolution effects on the uDiFFs at $z_1=0.1$ are much larger than
the corresponding ones at $z_1=0.5$.
At $z_1=0.1$, the uDiFFs of the $\pi^{+}\pi^{-}$ pair are much larger than
those of other hadron pairs.
The shapes of the curves of $z_2 D^{\pi^{+},\pi^{-}}_{q}$
in Fig.~(\ref{C}) are modified
by the QCD evolution. But their magnitudes remain almost the same compared
to the corresponding curves in Fig.~(\ref{A}).
This observation is also applicable to the case of the $\pi^{+}K^{-}$ pair,
except for the $s\to \pi^{+}K^{-}$ one in which the
value of $z_2D_{s}^{\pi^{+}K^{-}}$ becomes only half of that at $z_2=0.5$.\\

In the case of the $K^{+}K^{-}$ pair, the uDiFFS of NLChQM receive very little
QCD evolution effect. On the other hand, the QCD evolution changes
the uDiFFs in the NJL-jet model significantly both in the magnitudes and the shapes.
Furthermore, the uDiFFs of NLChQM here are much smaller than the ones of the NJL-jet model.\\

In the case of $z_1=0.5$ the magnitudes of NLChQM uDiFFs of the $\pi^{+}\pi^{-}$ pair are reduced,
but their $z_2$ dependencies remain more or less the same.
For the ones of the $\pi^{+}K^{-}$ pair, both of the magnitudes and the $z_2$ dependence of NLChQM uDiFFs are
very different from the ones at $z_1=0.1$. The magnitudes of
uDiFFs become one-fifth for the $u$ and $d$ quark, and one-tenth for the $s$ quark.
The NLChQM results of uDiFFs for the $K^{+}K^{-}$ pair are small when they are
compared with the other uDiFFs. Our last observation is the disfavored uDiFFs are all very small in the high $z_2$ regime in the both models.\\

\begin{figure}
\begin{tabular}{cc}
\includegraphics[width=8.0cm]{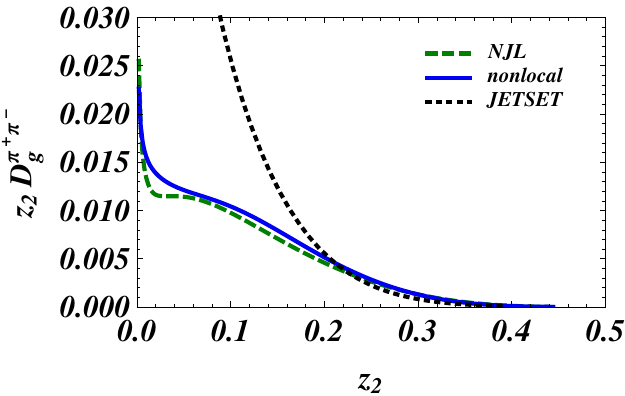}
\includegraphics[width=7.8cm]{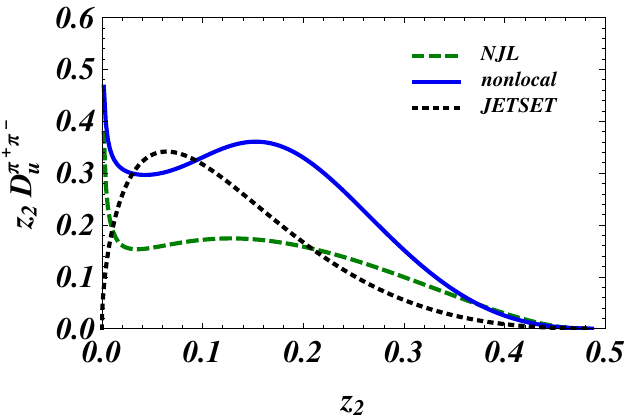}
\end{tabular}
\caption{Dihadron fragmentation functions of $z_2 D^{\pi^{+},\pi_{-}}_{g}(z_1,z_2)$ (left) and $z_2 D^{\pi^{+},\pi_{-}}_{u}(z_1,z_2)$ (right)
for $z_{1}=0.5$ at $Q^2=109\,\mathrm{GeV}^2$. The dashed, solid, and dotted lines denote the results of the NJL-jet model, the nonlocal chiral quark model,
and JETSET, respectively }
\label{G}
\end{figure}

The NLChQM and NJL-jet results for  $z_2 D^{\pi^{+}\pi^{-}}_{u}$ and $z_2 D^{\pi^{+}\pi^{-}}_{g}$ at
$Q^2=109\,\mathrm{GeV}^2$ are shown in Fig.~(\ref{G}).
We also present the results fitted from the output of the Monte Carlo event generator called JETSET \cite{JETSET}
by the following parametrization \cite{Majumder05}:
\begin{equation}
D(z_1,z_2)=Nz_{1}^{\alpha_1}z_2^{\alpha_2}(z_1+z_2)^{\alpha_3}(1-z_1)^{\beta_1}(1-z_2)^{\beta_2}(1-z_1-z_2)^{\beta_3}.
\end{equation}
where $N$, $\alpha_1$, $\alpha_2$,$\alpha_3$, $\beta_1$, $\beta_2$, and $\beta_3$ stand for
the parameters fitted by JETSET.

For the case of $g\to \pi^{+}\pi^{-}$,
the NLChQM result is
very similar to the one in the NJL-jet model. At the high-$z_2$ regime,
both the NLChQM and the NJL-jet results are close to the one obtained from
JETSET. On the other hand, the deviations appear at the low-$z_2$ regime.
For $u\to \pi^{+}\pi^{-}$, both the result of the NLCHQM and the NJL-jet model significantly differ from the one of JETSET
parametrization. The
shapes of the results of the NLChQM and the NJL-jet are similar, but the NLChQM one is larger in magnitude.
Both model results are divergent as $z_2$ approaches zero. The
shape of the result from JETSET is completely different, and it locates between the
NLChQM result and the NJL-jet model result.
\section{Summary and outlook}
In this article, we have investigated the unpolarized dihadron fragmentation functions (uDiFFs) of the pions and kaons
in the nonlocal chiral-quark model (NLChQM) and the NJL-jet model.
We find that the magnitudes of the curves
of $z_2 D^{h_1,h_2}_{q}(z_1,z_2)$ as a function of $z_2$ in the NLChQM and
the NJL-jet model are substantially different in every channels.
The difference between the results of these two models is more significant in the
small $z_2$ regime. In NLChQM, the uDiFFs of the $K^{+}K^{-}$ pair are particularly smaller than the uDiFFs of the other hadron pairs .
Moreover, they are also smaller compared with the corresponding ones in the NJL-jet model.
At $z_1=0.1$, the DiFFs of the $\pi^{+}\pi^{-}$ pair at $Q^2=4\,\mathrm{GeV}^2$ for the $u$, $d$ and $s$ quarks are comparable in their magnitudes. As $z_1$ increases to
$0.5$, $D_{u}^{\pi^{+}\pi^{-}}$ becomes larger than the other two DiFFs.\\

In the cases of the $\pi^{+}K^{-}$ pair, the uDiFF for the $s$ quark at
$Q^2=4\,\mathrm{GeV}^2$ is dominant over the other two, for both cases of $z_1=0.1$ and $0.5$. At $Q^2=109\,\mathrm{GeV}^2$, we observe that the DiFFs of
$g\to \pi^{+}\pi^{-}$ in the NJL-jet model and NLChQM are very similar,
but both are smaller than the JETSET result in the small-$z$ region.
On the contrary, the NLChQM uDiFF of $u \to \pi^{+}\pi^{-}$ is larger
than the JETSET result which is larger than the NJL-jet result.
Nevertheless, the $z$-dependencies of the uDiFFs in the NLChQM and the NJL-jet
model are similar. But they are very different from the one of JETSET.
The disfavored uDiFFs are all suppressed at high $z_2$ regime in both models.\\
In summary, we have applied the NLChQM to study the uDiFFs of the pions and kaons. Our next step is to include the vector mesons in our model and extend our study to the Collins fragmentation functions and
the polarized DiFFs $H^{\sphericalangle, h_{1},h_{2}}_{q}$ which are important in the extraction of the transversity parton distribution of the hadrons.\\

\section*{Acknowledgments}
S.i.N. is very grateful to the hospitality during his visiting National Taiwan University (NTU) with the financial support from NCTS (North) of Taiwan, where
the present work was partially performed. C.W.K. are supported by the grant NSC 102-2112-M-033-005-MY3 from National Science Council (NSC) of Taiwan. C.W.K. are also supported by the grants  MOST 105-2112-M-033-004, MOST 106-2112-M-033-003,MOST 107-2119-M-033-002 and MOST 108-2112-M-033-004 and MOST 109-2112-M-033-002.
F.J.J. and D.J.Y are partially supported by NSC of Taiwan (grant No. 102-2112-M-003-004-MY3).
The work of S.i.N. was supported by the National Research Foundation of Korea (NRF) grants, funded by the Korean government (MSIT) (No. 2018R1A5A1025563 and 2019R1A2C1005697).
\section*{Appendix}
In this appendix, we would like to explain why the results in Ref.~\cite{Casey:2012ux} are incorrect.
For completeness, the relevant equations are listed here. The DiFFs are given by
\begin{eqnarray}
D^{h_{1},h_{2}}_{q}(z_{1},z_{2})&=&\delta_{aq}\hat{d}^{h_{1}}_{q}(z_{1})\frac{D^{h_{2}}_{q_{1}}\left(\frac{z_{2}}{1-z_{1}}\right)}{1-z_{1}}+
\delta_{bq}\hat{d}^{h_{2}}_{q}(z_{2})\frac{D^{h_{1}}_{q_{2}}\left(\frac{z_{1}}{1-z_{2}}\right)}{1-z_{2}} \nonumber \\
&+&\sum_{Q}\int^{1}_{z_{1}+z_{2}}\frac{d\eta}{\eta^2}\hat{d}^{Q}_{q}(\eta)D^{h_{1},h_{2}}_{Q}\left(\frac{z_{1}}{\eta},\frac{z_2}{\eta}\right).
\label{appDihadron1}
\end{eqnarray}
The flavor contents of $h_{1}$ and $h_2$ are $h_1=(a\bar{q}_{1})$ and $h_{2}=b\bar{q}_{2}$.
To simplify the equation we make change of valuables $\xi_{1}=z_{1}/\eta$ and $\xi_{2}=z_2/\eta$:
\begin{eqnarray}
D^{h_{1},h_{2}}_{q}(z_{1},z_{2})&=&\delta_{qa}\hat{d}^{h_{1}}_{q}(z_{1})\frac{D^{h_{2}}_{q_{1}}\left(\frac{z_{2}}{1-z_{1}}\right)}{1-z_{1}}+
\delta_{qb}\hat{d}^{h_{2}}_{q}(z_{2})\frac{D^{h_{1}}_{q_{2}}\left(\frac{z_{1}}{1-z_{2}}\right)}{1-z_{2}} \nonumber \\
&+&\sum_{Q}\int^{\frac{z_1}{z_{1}+z_{2}}}_{z_{1}}d\xi_{1}\int^{\frac{z_1}{z_1+z_2}}_{z_2}d\xi_{2}\delta(z_2\xi_{1}-z_{1}\xi_{2})
\hat{d}^{Q}_{q}(z_{1}/\xi_{1})D^{h_{1},h_{2}}_{Q}(\xi_{1},\xi_{2}).
\label{appDihadron2}
\end{eqnarray}

After evaluating the delta function, the third term of Eq.~(\ref{appDihadron2}) is simplified as
\begin{eqnarray}
\label{appdis0}
&&\sum_{Q}\int^{\frac{z_1}{z_{1}+z_{2}}}_{z_{1}}d\xi_{1}\int^{\frac{z_1}{z_1+z_2}}_{z_2}d\xi_{2}\delta(z_2\xi_{1}-z_{1}\xi_{2})
\hat{d}^{Q}_{q}(z_{1}/\xi_{1})D^{h_{1},h_{2}}_{Q}(\xi_{1},\xi_{2}) \nonumber \\
&&=\sum_{Q}\int^{\frac{z_1}{z_{1}+z_{2}}}_{z_{1}}\frac{d\xi_{1}}{z_1}
\hat{d}^{Q}_{q}(z_{1}/\xi_{1})D^{h_{1},h_{2}}_{Q}\left(\xi_{1},\frac{z_2\xi_{1}}{z_1}\right).
\end{eqnarray}

In our numerical calculation, the values of $D^{h_1,h_2}_{q}(z_1,z_2)$ are defined at the
grid points:
\begin{equation}
z_1=I\Delta\xi,\,\, z_2=J\Delta\xi,\,\,\Delta \xi=1/N_\mathrm{grid}. \,\,\,\,\,\,\,\,\,
I,J=1,2,3....N_\mathrm{grid}.
\label{grid}
\end{equation}
The integral in the RHS of Eq.~\ref{appdis0} is discretized as the Riemann sum:
\begin{eqnarray}
&&\sum_{Q}\sum_{A=I}^{M}\frac{\Delta\xi}{z_1}
\hat{d}^{Q}_{q}\left(\frac{z_{1}}{A\Delta\xi}\right)D^{h_{1},h_{2}}_{Q}\left(A\Delta\xi,\frac{z_2}{z_1}A\Delta\xi\right) \nonumber \\
&=&\sum_{Q}\frac{1}{N_\mathrm{grid}}\sum_{A=I}^{M}\frac{1}{z_1}
\hat{d}^{Q}_{q}\left(\frac{z_{1}}{\frac{A}{N_\mathrm{grid}}}\right)D^{h_{1},h_{2}}_{Q}\left(\frac{A}{N_\mathrm{grid}},\frac{z_2}{z_1}\frac{A}{N_\mathrm{grid}}\right).
\label{discretization}
\end{eqnarray}
Here $z_1=\frac{I}{N_\mathrm{grid}}$ and $\frac{M}{N_\mathrm{grid}}<\frac{z_1}{z_1+z_2}<\frac{M+1}{N_\mathrm{grid}}$.
Moreover, if the arguments of the functions are not at the grid points then we determine their values by interpolation.
For example, if $\frac{K}{N_\mathrm{grid}}<\zeta<\frac{K+1}{N_\mathrm{grid}}$ ($K$ is an integer) then
\begin{equation}
\hat{d}^{h}_{Q}(\zeta)\approx (K+1-N_\mathrm{grid}\zeta)\hat{d}^{h}_{Q}\left(\frac{K}{N_\mathrm{grid}}\right)
+(N_\mathrm{grid}\zeta-K)\hat{d}^{h}_{Q}\left(\frac{K+1}{N_\mathrm{grid}}\right).
\end{equation}
Similarly, if $\frac{K}{N_\mathrm{grid}}<\zeta_1<\frac{K+1}{N_\mathrm{grid}}$ and $\frac{L}{N_\mathrm{grid}}<\zeta_2<\frac{L+1}{N_\mathrm{grid}}$
($K$ and $L$ are both integers)
then we have
\begin{eqnarray}
D^{h_1,h_2}_{Q}(\zeta_1,\zeta_2)&\approx&(K+1-N_\mathrm{grid}\zeta_1)(L+1-N_\mathrm{grid}\zeta_2)D^{h_1,h_2}_{Q}
\left(\frac{K}{N_\mathrm{grid}},\frac{L}{N_\mathrm{grid}}\right) \nonumber \\
&+&(K+1-N_\mathrm{grid}\zeta_1)(N_\mathrm{grid}\zeta_2-L)D^{h_1,h_2}_{Q}\left(\frac{K}{N_\mathrm{grid}},\frac{L+1}{N_\mathrm{grid}}\right) \nonumber \\
&+&(N_\mathrm{grid}\zeta_1-K)(L+1-N_\mathrm{grid}\zeta_2)D^{h_1,h_2}_{Q}\left(\frac{K+1}{N_\mathrm{grid}},\frac{L}{N_\mathrm{grid}}\right) \nonumber \\
&+&(N_\mathrm{grid}\zeta_1-K)(N_\mathrm{grid}\zeta_2-L)D^{h_1,h_2}_{Q}\left(\frac{K+1}{N_\mathrm{grid}},\frac{L+1}{N_\mathrm{grid}}\right).
\end{eqnarray}
The summation in Eq.~(\ref{discretization}) converges when $N_\mathrm{grid}$ is large enough.
Applying a Gauss-Seidel-type iteration method to both Eqs~(\ref{appDihadron1}) and (\ref{appDihadron2}), we arrive at
results which are significantly different from those obtained in Ref.~\cite{Casey:2012ux}.

Interestingly, we are able to
reproduce the results in Ref.~\cite{Casey:2012ux} by inserting an extra factor $\frac{1}{200}$ in the third term of Eqs.~(\ref{appDihadron1}) and (\ref{appDihadron2}). In other words, instead of solving Eq.~(\ref{appDihadron1}),
the authors of Ref.~\cite{Casey:2012ux} actually have solved the following equation:
\begin{eqnarray}
D^{h_{1},h_{2}}_{q}(z_{1},z_{2})&=&\delta_{aq}\hat{d}^{h_{1}}_{q}(z_{1})\frac{D^{h_{2}}_{q_{1}}\left(\frac{z_{2}}{1-z_{1}}\right)}{1-z_{1}}+
\delta_{bq}\hat{d}^{h_{2}}_{q}(z_{2})\frac{D^{h_{1}}_{q_{2}}\left(\frac{z_{1}}{1-z_{2}}\right)}{1-z_{2}} \nonumber \\
&+&\frac{1}{200}\sum_{Q}\int^{1}_{z_{1}+z_{2}}\frac{d\eta}{\eta^2}\hat{d}^{Q}_{q}(\eta)D^{h_{1},h_{2}}_{Q}\left(\frac{z_{1}}{\eta},\frac{z_2}{\eta}\right).
\label{appDihadron3}
\end{eqnarray}

\begin{figure}[t]
\begin{tabular}{cc}
\includegraphics[width=7.2cm]{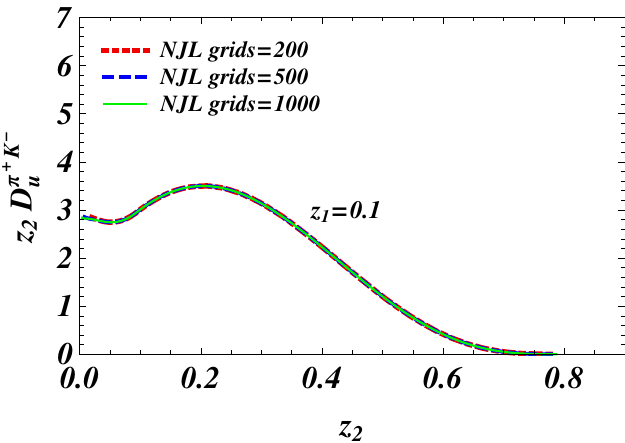}
\includegraphics[width=7.2cm]{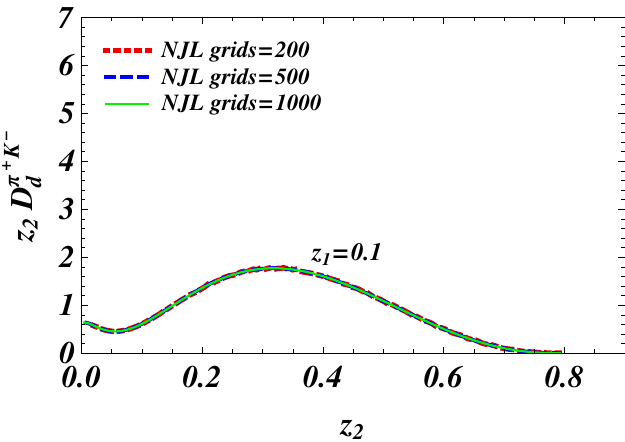}
\end{tabular}
\begin{tabular}{cc}
\includegraphics[width=7.2cm]{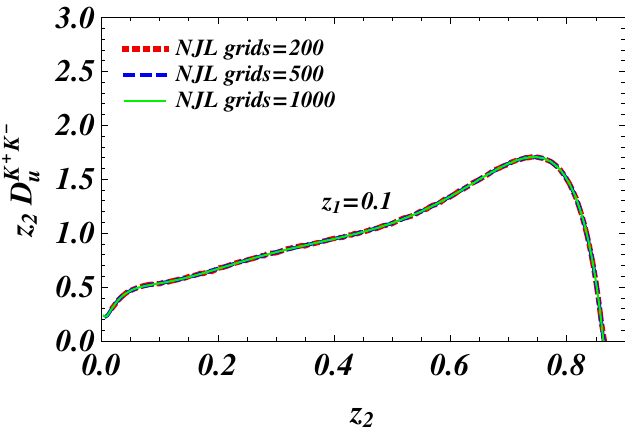}
\includegraphics[width=7.2cm]{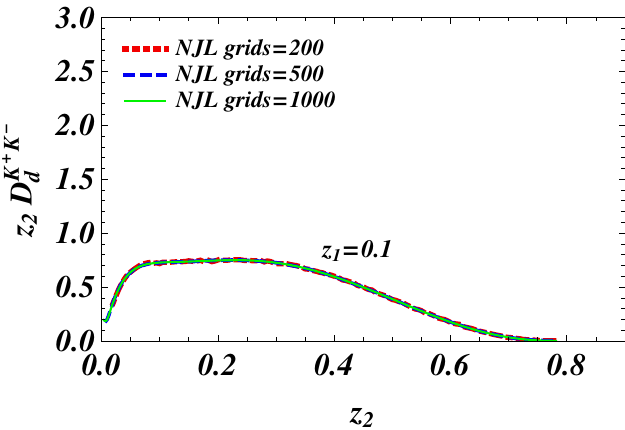}
\end{tabular}
\begin{tabular}{cc}
\includegraphics[width=7.2cm]{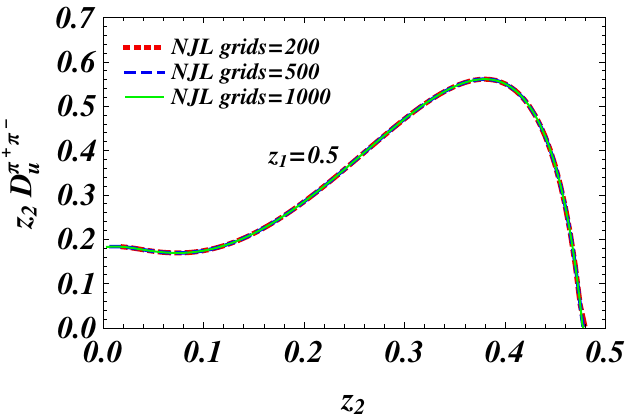}
\includegraphics[width=7.2cm]{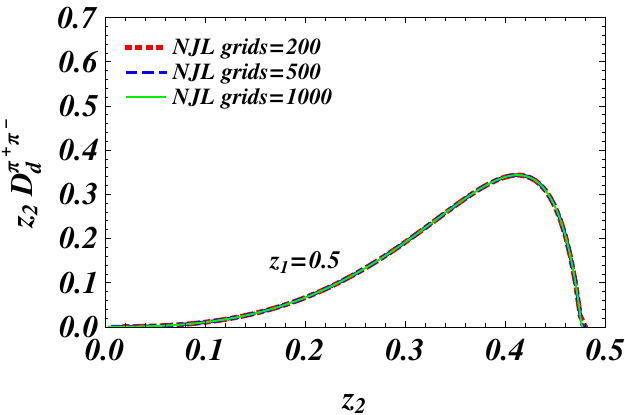}
\end{tabular}
\caption{Some of the Dihadron fragmentation functions in the NJL-jet
model obtained by solving Eq.~(\ref{appDihadron3}).}
\label{appA}
\end{figure}

We have solved Eq.~(\ref{appDihadron3}) with $N_\mathrm{grid}$=200,~500, and 1000.
Some results are depicted in Fig.~(\ref{appA}). It is obvious that we obtain the same results presented in Ref. ~\cite{Casey:2012ux} and our results converge as $N_\mathrm{grid}\ge 200$.

Where is this factor $1/200$ from? Note that in Ref.~\cite{Casey:2012ux} their $N_\mathrm{grid}$ is chosen to be 200.
Hence, we suspect that the authors in Ref.~\cite{Casey:2012ux} have accidently introduced an extra factor
$1/N_\mathrm{grid}$ in the discretization of the third term in Eq.~(\ref{appDihadron2}) when evaluating the delta function.
Namely we believe that the authors in Ref.~\cite{Casey:2012ux} actually have solved the following equation:

\begin{eqnarray}
D^{h_{1},h_{2}}_{q}(z_{1},z_{2})&=&\delta_{aq}\hat{d}^{h_{1}}_{q}(z_{1})\frac{D^{h_{2}}_{q_{1}}\left(\frac{z_{2}}{1-z_{1}}\right)}{1-z_{1}}+
\delta_{bq}\hat{d}^{h_{2}}_{q}(z_{2})\frac{D^{h_{1}}_{q_{2}}\left(\frac{z_{1}}{1-z_{2}}\right)}{1-z_{2}} \nonumber \\
&+&\frac{1}{N_\mathrm{grid}}\sum_{Q}\int^{1}_{z_{1}+z_{2}}\frac{d\eta}{\eta^2}\hat{d}^{Q}_{q}(\eta)D^{h_{1},h_{2}}_{Q}\left(\frac{z_{1}}{\eta},\frac{z_2}{\eta}\right).
\label{appDihadron4}
\end{eqnarray}

We have solved Eq.~(\ref{appDihadron4}) with $N_\mathrm{grid}$=50,~100,~200,~500 and 1000.
We present our result of
$z_2D_u^{\pi^+\pi^{-}}(z_1,z_2)$ with $z_1 = 0.1$
in Fig.~(\ref{appB}). It agrees excellently with the corresponding figure in Ref.~\cite{Casey:2012gi} by the
same authors of Ref.~\cite{Casey:2012ux}.
This provides
a convincing evidence that indeed the authors of Ref.~\cite{Casey:2012ux} have actually
solved Eq.~(\ref{appDihadron4}).

Furthermore, we also demonstrate our results of uDiFFs of other pairs
in Fig.~({\ref{appC}). They are the solutions of the equation Eq.~({\ref{appDihadron4}).
In contrast to the case of $z_2D_u^{\pi^+\pi^{-}}(z_1,z_2)$ with $z_1 = 0.1$,
Our results indicate that some results do not converge even for $N_\mathrm{grid} = 1000$.
This is in contradiction to the claim reported in Ref.~\cite{Casey:2012ux} that the results converge to within $5$
percent for $N = 200$.

\begin{figure}
\includegraphics[width=10.0cm]{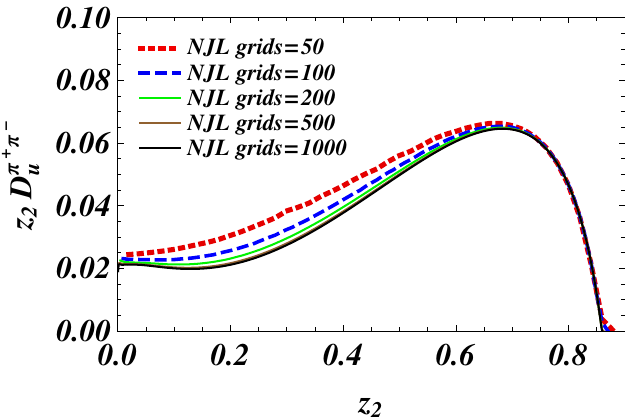}
\caption{$z_2D_u^{\pi^{+}\pi^{-}}(z_1,z_2)$ with $z_1 = 0.1$ obtained by solving Eq.~(\ref{appDihadron4}).}
\label{appB}
\end{figure}

\begin{figure}[t]
\begin{tabular}{ccc}
\includegraphics[width=7.2cm]{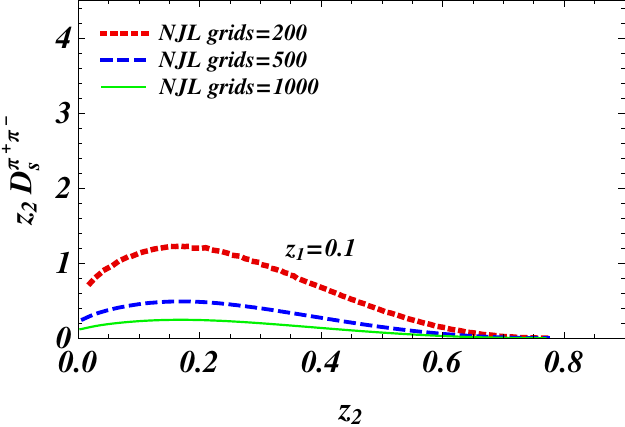}
\includegraphics[width=7.2cm]{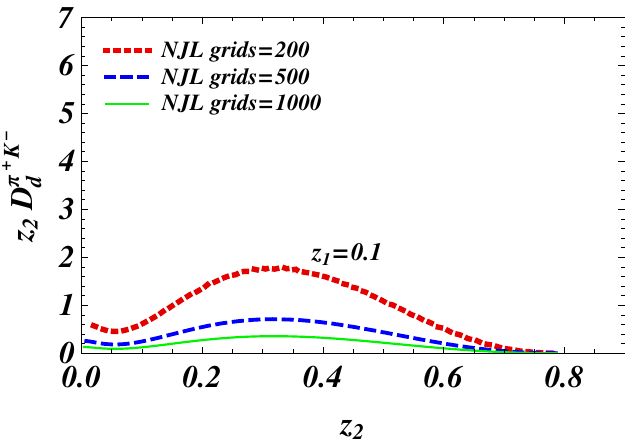}
\end{tabular}
\begin{tabular}{ccc}
\includegraphics[width=7.2cm]{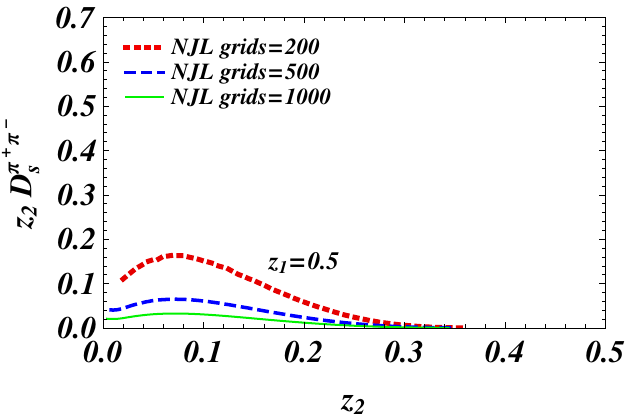}
\includegraphics[width=7.2cm]{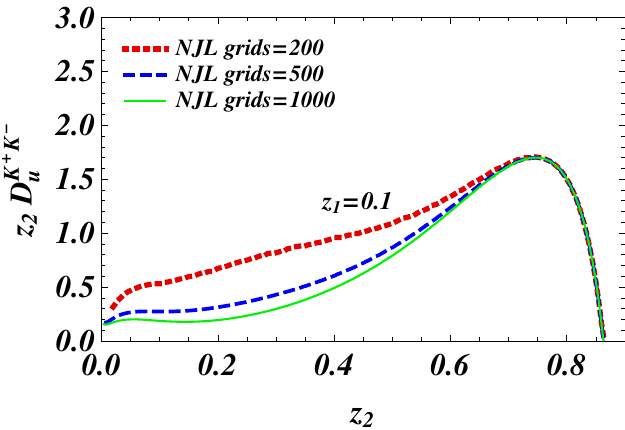}
\end{tabular}
\caption{Some Dihadron fragmentation functions in the NJL-jet
model as the solutions of Eq.~(\ref{appDihadron4}).}
\label{appC}
\end{figure}

In conclusion, for the NJL-jet model, we solve the related integral equations Eqs.~(\ref{appDihadron1})
and (\ref{appDihadron2}) directly using an iteration method and reach the results of
DiFFs which are significantly different from the ones in Ref.~\cite{Casey:2012ux}.
We have shown that it is very likely the calculations done in Ref.~\cite{Casey:2012ux}
actually solve the incorrect equations with an extra factor. It would be
useful to clarify the discrepancy
if the authors of Ref.~\cite{Casey:2012ux} perform the similar investigation.


\end{document}